\documentclass[aps,prc,twocolumn,floatfix]{revtex4}
\usepackage[dvips]{graphicx}
\usepackage{psfig}
\usepackage{bm}

\newcommand{\simleq}{\; \raisebox{-0.4ex}{\tiny$\stackrel
{{\textstyle<}}{\sim}$}\;}
\newcommand{\up}[2]{$^{#1#2}$}
\newcommand{\upt}[3]{$^{#1#2#3}$}

\newcommand{\BibTitle}[1]{}
\newcommand{\doetal}[1]{\ {\em et al.\/}}
\newcommand{\parp}{$^+$}

\begin{document}
\title{Gamow-Teller $GT_+$ distributions in nuclei with mass $A=90-97$} 
\author{A.\ Juodagalvis$^{1,2}$ and D.J.\ Dean$^1$}
\affiliation{$^1$Physics Division, Oak Ridge National Laboratory,
P.O.\ Box 2008, Oak Ridge, TN 37831-6373} 
\affiliation{$^2$Department of Physics and Astronomy, University of
Tennessee, Knoxville, TN 37996}
\date{June 17, 2004}
\begin{abstract}
We investigate the Gamow-Teller strength distributions in the
electron-capture direction in nuclei having mass $A=90-97$, assuming a
$^{88}$Sr core and using a realistic interaction that reasonably
reproduces nuclear excitation spectra for a wide range of nuclei in the
region as well as experimental data on Gamow-Teller strength
distributions.  We discuss the systematics of the distributions and
their centroids.  We also predict the strength distributions for
several nuclei involving stable isotopes that should be experimentally
accessible for one-particle exchange reactions in the near future.
\end{abstract}

\maketitle

\section{Introduction}
\label{sect-Intro}

New frontiers of nuclear structure experiments to probe the
Gamow-Teller distributions in medium-mass nuclei are currently being
pursued. These experiments will be able to measure Gamow-Teller data
in the mass 90-100 region.  Extensive theoretical studies have been
devoted to Gamow-Teller total strengths and strength distributions in
$1s$-$0d$ shell nuclei (mass $A=16$-40 nuclei) \cite{BW-88} and the
$0f$-$1p$ shell (mass $A=40$-80 nuclei)
\cite{Langanke95,Martinez96,Caurier99GT}.  Due to an excellent
agreement between shell model results and the available experimental data,
the calculated results have been used extensively to predict numerous
Gamow-Teller strength distributions in nuclei that have not yet become
experimentally accessible \cite{Langanke00}.

In addition to their nuclear structure interests, an appropriate
description of Gamow-Teller transitions in nuclei directly affects the
early phases of type II supernova core collapse since electron capture
rates are partly determined by them.  The effects of the improved rate
estimates are rather dramatic, as was recently discussed in
Refs.~\cite{Langanke03,Hix03}.  In addition to the standard
Gamow-Teller transitions, first- and second-forbidden transitions
contribute to the electron capture rates in the supernova environment.
For terrestrial experiments, the primary focus is on the Gamow-Teller
transitions.

Recently, Zegers {\em et al\/} \cite{Zegers04} proposed to measure the
Gamow-Teller distributions using stable Zr and Mo isotopes as targets
in $(t,{}^3{\rm He})$ reactions \cite{Mo-Zegers}.  Estimates indicate
that the Gamow-Teller strength is sufficiently large to be measured.
In this paper, we will investigate these transitions using standard
shell-model diagonalization techniques for 36 nuclei with the mass
number $90\leq A\leq97$ ($Z=40$-47, $N=50$-57). To validate the
interaction, we also studied excitation spectra in those and other
nuclei in the region. Since our model space does not contain all
spin-orbit partners, i.e.\ it is not a complete $0\hbar\omega$
calculation, the total Gamow-Teller strength will be
overestimated in our calculations. We adopt a single
quenching factor similar to the one discussed in Ref.\
\cite{N50GT-Brown}.  We estimated this factor based on recent
experimental data on \up97Ag \cite{Ag97GT-Hu}. 
We used this measurement to gauge our calculation for two reasons.
First, it used the total absorption spectrometry which accounts also for
the weak $\gamma$-ray cascades that follow the $\beta^+$ decay. 
Second, almost all total Gamow-Teller
strength is inside the $Q$-window. We note that this factor need not
be universal as it is simply a phenomenological tool at this point.

The remainder of this paper is organized as follows. In Section
\ref{sect-Model}, we present results on the nuclear spectra, generated
with an effective interaction that uses $^{88}$Sr as a core, and
compare them to experiment.  In section \ref{sect-GT}, we present our
shell-model diagonalization results for the Gamow-Teller strength
distributions and compare to experiment when available. We also
present systematics of the Gamow-Teller centroids.  In section
\ref{sect-SMCode}, we discuss the distributed-memory shell-model
computer code that we developed and used for these calculations.
Finally, we conclude and give a perspective in Section
\ref{sect-Summary}.

\section{Calculated spectra using the $^{\bm8\bm8}$Sr core}
\label{sect-Model}

We perform our shell-model diagonalization calculations
in a model space taking $^{88}$Sr as the core nucleus and allowing
excitations within the valence space 
of $1p_{1/2}$ and $0g_{9/2}$ proton shells and $1d_{5/2}$,
$2s_{1/2}$, $1d_{3/2}$, $0g_{7/2}$, and $0h_{11/2}$ neutron shells.  
While it cannot be used for 
calculations of $\beta$-decays, it appears suitable for
Gamow-Teller distributions in the electron-capture direction.
The effective interaction \cite{In102-Lipoglavsek} was derived from a CD-Bonn potential
\cite{CDBonn} using the machinery of many-body perturbation theory 
\cite{HjorthJensen95}. We use
the following single-particle energies: 
$\varepsilon(p_{1/2})=0.0$ MeV and
$\varepsilon(g_{9/2})=0.9$ MeV for protons; and
$\varepsilon(d_{5/2})=0.0$ MeV, $\varepsilon(s_{1/2})=1.26$ MeV,
$\varepsilon(d_{3/2})=2.23$ MeV, $\varepsilon(g_{7/2})=2.90$ MeV, and
$\varepsilon(h_{11/2})=3.50$ MeV for neutrons. 
A slightly different version of this interaction was used to describe Sr
and Zr isotopes \cite{SrZr-Holt}. We have not attempted to adjust
the interaction to obtain a better fit to experimental data \cite{NDSOnline}.

\begin{figure*}[tbp]
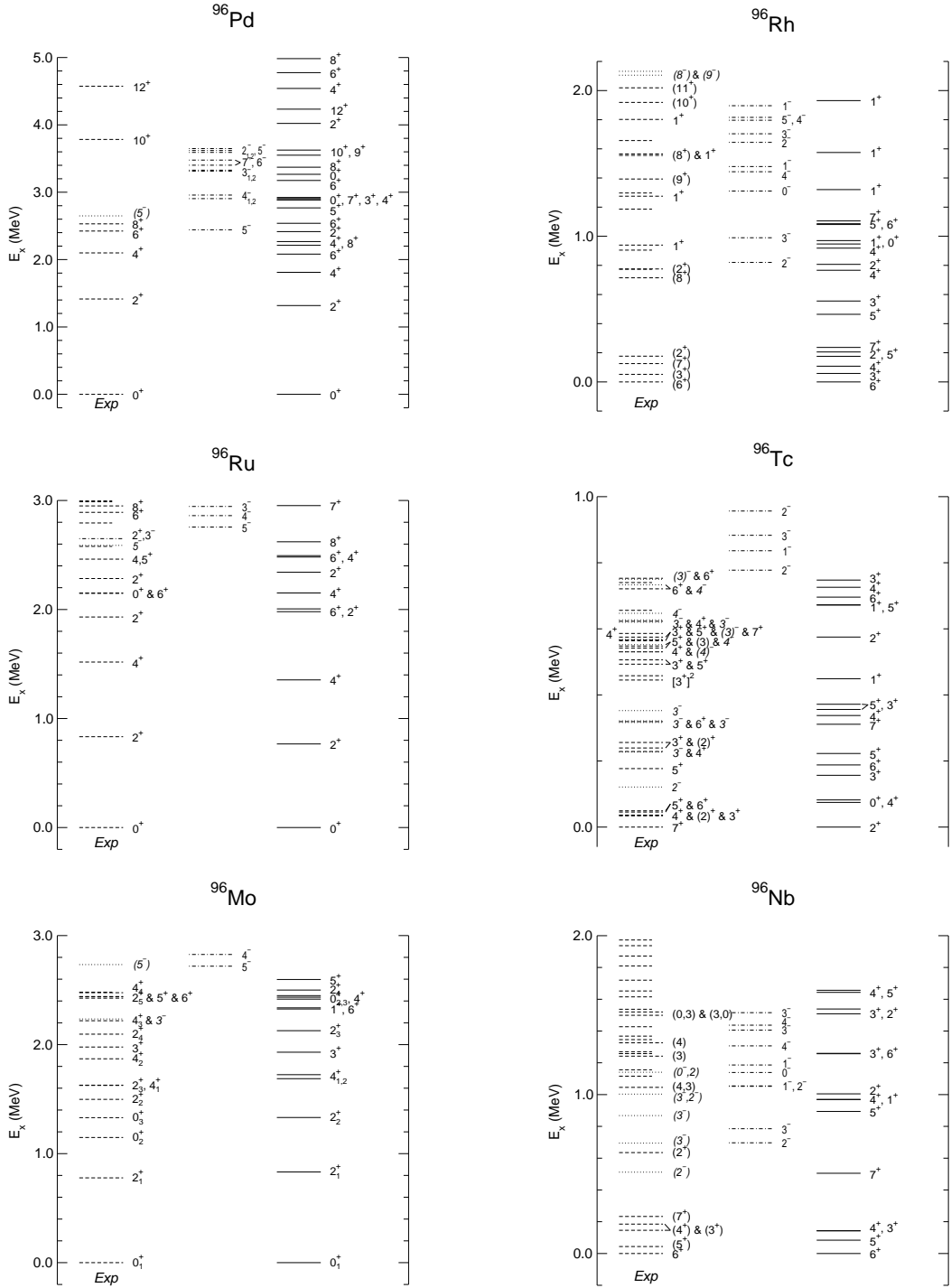

\centerline{
\hfill
  \psfig{figure=figPd96-Spectrum.eps,width=0.33\textwidth,angle=270}
  \hfill
  \psfig{figure=figRh96-Spectrum.eps,width=0.33\textwidth,angle=270}
\hfill
}
\vspace{4mm}
\centerline{
\hfill
  \psfig{figure=figRu96-Spectrum.eps,width=0.33\textwidth,angle=270}
  \hfill
  \psfig{figure=figTc96-Spectrum.eps,width=0.33\textwidth,angle=270}
\hfill
}
\vspace{4mm}
\centerline{
\hfill
  \psfig{figure=figMo96-Spectrum.eps,width=0.33\textwidth,angle=270}
  \hfill
  \psfig{figure=figNb96-Spectrum.eps,width=0.33\textwidth,angle=270}
\hfill
}
\caption{\label{fig-A96Spectra} 
Experimental \cite{A96DataSheets} and calculated spectra in
\up96Pd, \up96Rh, \up96Ru, \up96Tc, \up96Mo, and \up96Nb.
Experimental levels are shown in the left column; the middle and right
columns display calculated negative and positive parity states,
respectively. Experimental positive parity states are shown using
dashed lines; negative parity states are shown using dotted lines;
states with known spin and uncertain parity are shown using
dash-dotted lines; the states having undetermined both spin and parity
are shown using a shorter dashed line. The sign '\&' indicates that
assignments refer to different experimental states, with the second
one lying at a higher excitation energy. Since there is no
spin ambiguity in the calculation, a comma is used in the middle and
right columns in a similar situation.
In some cases a subscript number is used to distinguish levels
having the same spin and parity.
At most, 10 calculated negative parity states are shown for each nucleus.
Only 1\parp\ calculated states are shown in \up96Rh above 
$E_x\approx1.1$ MeV. Only negative parity states and positive parity
states with $J>5$ are 
shown for \up96Ru above $E_x=2.5$ MeV in experimental spectrum.
Experimental positive parity states are cut-off \up96Mo after
$E_x=2.5$ MeV. 
}
\end{figure*}
\begin{figure}[tb]
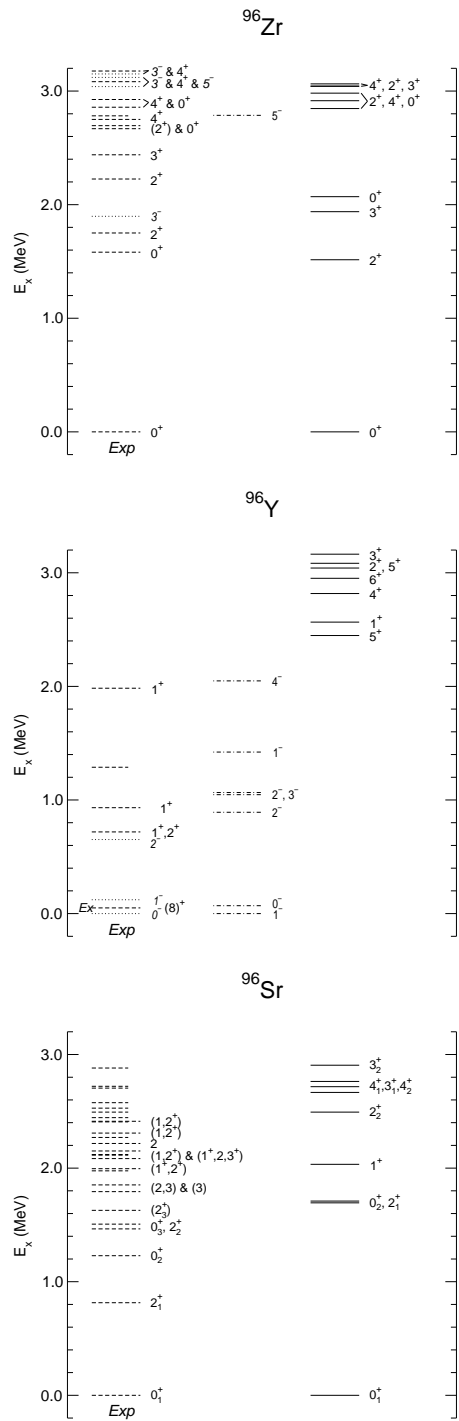

\centerline{
\psfig{figure=figZr96-Spectrum.eps,width=0.33\textwidth,angle=270}
}
\vspace{4mm}
\centerline{
\psfig{figure=figY96-Spectrum.eps,width=0.33\textwidth,angle=270}
}
\vspace{4mm}
\centerline{
\psfig{figure=figSr96-Spectrum.eps,width=0.33\textwidth,angle=270}
}
\caption{
\label{fig-A96Spectra2} Experimental and calculated spectra in
\up96Zr, \up96Y, and \up96Sr. For notations, see caption of Fig.\
\ref{fig-A96Spectra}.
Excitation energy of the isomeric state in \up96Y with $J^\pi=8^+$
is not known \cite{A96DataSheets}. 
}
\end{figure}

We calculated low-energy spectra of more than 50 nuclei with masses
$90\leq A\leq 98$, $38\leq Z\leq48$, and $50\leq N\leq 58$.
General agreement between the calculated lowest states and
experimentally observed states is satisfactory. We judged the
agreement based on reproduction of low-lying states up to a chosen
excitation energy.  For odd-odd isotopes, which have higher density of
states, the upper limit was chosen to be 1 MeV. For even-even nuclei
the limit was up to 3 MeV. Not all observed states were found in the
model space, as would be expected from a restricted calculation, and
for some nuclei our calculations suggested some low-lying states that
have not yet been observed.

The interaction generally reproduces the correct spin for the lowest
states of both parities as well as their ordering,
though there are cases where some levels are interchanged.  The energy
splitting between the lowest states with different parities is
reproduced with varying success, although this is difficult to judge for
some nuclei because of the lack of experimental information. 

For representative spectra which indicate the overall quality of the
interaction, we show nuclei having mass $A=96$ in Figs.\
\ref{fig-A96Spectra}-\ref{fig-A96Spectra2}.  The maximum energy range
shown in the plots is varied following the density of states.  From these
figures we observe that the spectra of even-even nuclei (\up96Pd,
\up96Ru, \up96Mo, \up96Zr, and \up96Sr) are reproduced well. For
\up96Pd there are more calculated states than experimentally
known. In some nuclei the model space is insufficient to
describe all observed states. Odd-odd nuclei (\up96Rh,\up96Tc,
\up96Nb, and \up96Y), having more states, are also more difficult to
describe although, even here, the interaction performs reasonably well. 
The position of 8\parp\ isomer in \up96Y is not known experimentally
\cite{A96DataSheets}. This state appears in the
calculation at a relatively high excitation energy, 1.1 MeV above the
lowest positive parity state which was calculated to be 5\parp.  The
nucleus \up96Tc reflects a situation where the lowest states are
experimentally very close (in this nucleus there are 6 states in the
energy range of 50 keV), while the calculation reproduces the states
but not their energies (the calculated range is 310 keV). A similar
situation occurs in $^{92,94}$Nb.

These $A=96$ nuclei reflect the situation in other cases as well, with
a general conclusion that the interaction reproduces excitation
spectra reasonably well, though fine tuning might increase the
accuracy. We do not discuss them in a greater detail, since the focus
of our paper is Gamow-Teller distributions.

\section{Gamow-Teller strength distributions}
\label{sect-GT}


Our study focuses on the Gamow-Teller transitions from the
lowest positive parity states which is natural for the most
nuclei in the region above \up88Sr, with the exception of the 
Y isotopes where the odd proton in the
$p_{1/2}$ shell is responsible for low-lying negative parity states. 
Since our model space is not sufficient to reasonably
reproduce negative parity states in Sr isotopes where no valence
protons are available, we do not calculate the transitions between
these two isotope chains.  
Among the calculated nuclei, there are
three cases where we chose the lowest 
experimental state to be the initial state for 
$GT$ excitations rather than using our calculated lowest 
energy state.  This affected two $N=51$ nuclei,
\up92Nb and \up94Tc, where the calculation places 2\parp\ to be the
lowest state, and the nucleus \up96Tc.

The Gamow-Teller strength was calculated using the formula
\begin{equation}
GT_+\, =\, \langle \bm{\sigma } \bm{\tau} \rangle^2\,
 =\, 
 \frac1{2J_i+1} \sum\limits_f 
   \left| \langle \Psi_f ||
   \sum_k \bm{\sigma}(k) \bm{\tau_+}(k) || \Psi_i \rangle \right|^2.
\end{equation}
To obtain the strength distribution, we used the method of moments
\cite{StrengthFunction}.  
We performed 33 iterations for each $J_f$ in all nuclei except for the
decays of \up97Mo, where we did 24 iterations per final
state, and \up97Ag, where a complete convergence was achieved.
The $GT_+$ strength inside the experimental
$Q_{EC}$ window \cite{Audi03} is marked as $B_{Q_{EC}}$. This value is
only an estimate, since we did not strive to achieve the convergence
of states inside the $Q$-window.

As discussed above, our calculation overpredicts the Gamow-Teller
strength; thus we include a hindrance factor, $h$, so that
$S(GT_+)=GT_+/h$.  This factor is found by comparing
experimental data to the calculated Gamow-Teller total strength. For
nuclei around \upt100Sn, the single-particle estimate of the
Gamow-Teller strength is commonly used, since the main contribution
comes from a transition of a
$g_{9/2}$ proton into a $g_{7/2}$ neutron.
The estimate is given by a formula (see e.g.\ \cite{Towner85}):
\begin{equation}
  \sum GT_+= 
  \frac{N_{9/2}}{10}\, 
  \left(1-\frac{N_{7/2}}8\right)\,
  GT_+(^{100}{\rm Sn}),
\label{eq-GTocc}
\end{equation}
where $N_{9/2}$ is the occupation of the $g_{9/2}$ shell by protons,
and $N_{7/2}$ is the occupation of $g_{7/2}$ shell by neutrons in the
initial state of a parent nucleus, and $GT_+(^{100}{\rm Sn})=17.78$ is
the single particle estimate of the total Gamow-Teller strength for
\upt100Sn.  In the simplest, non-interacting shell model, the occupation
numbers are replaced by the numbers of valence particles in the
corresponding shells.  This simplest estimate does not exactly
reproduce our calculated strength even though the values are close. It
underestimates the strength for isotopes below the mass $A=96$, which
is not surprising because our model space allows excitations out of
the $p_{1/2}$ shell. On the other hand, it overestimates the strength for
some isotopes with $A=97$, which is related to the 
partial occupation of the $g_{7/2}$ shell by neutrons reducing the
total strength as compared to the non-interacting picture.

Since our calculated total strength is reasonably close to the
single-particle estimates, we could use the experimental hindrance
factor quoted relative to the single-particle estimate:
$h^{exp}=GT_{sp}/GT_{exp}$.  Unfortunately, in this region,
experimental information on $h^{exp}$ is limited.  Some of the
calculated nuclei naturally decay from the ground state by
$\beta^-$-decay instead of electron capture, while in other nuclei,
the $Q$-window contains only a small fraction of the total strength.
Thus the total $GT_+$ strength could be obtained only by $(n,p)$ or
similar one-particle exchange reactions.
An additional uncertainty in deriving the hindrance factor, even for
nuclei where the $Q$-window is large, comes from the fact
that $\gamma$-ray spectroscopy misses a significant fraction of the
Gamow-Teller decay strength due to sensitivity limits of detectors, a
low population of nuclear levels close to the $Q$-limit as well as
weak intensity of their decays \cite{N50GT-Brown,In102GT-Gierlik}.
This limitation can be largely overcome combining a high-resolution
$\gamma$-ray detector with total absorption spectrometry (TAS), as
was done in a number of recent experiments on nuclei in the \upt100Sn
region. For example, a study of \up97Ag decay \cite{Ag97GT-Hu} showed
that only 2/3 of the total Gamow-Teller strength is obtained by
high-resolution $\gamma$-ray spectrometry, while the same number for
\upt102In was about 1/8 \cite{In102GT-Gierlik}.

One nucleus, \up97Ag,
has almost 98\% of the total $GT$ strength inside the $Q_{EC}$
window. We can use this nucleus to estimate the experimental hindrance
factor. Hu {\em et al\/} \cite{Ag97GT-Hu} reported $\sum
B(GT)=3.00(40)$ based on TAS measurements, which leads to the
hindrance factor $h^{exp}=4.24_{-0.50}^{+0.65}$. Another nucleus where
this window is large, and there is a TAS measurement available is
\up98Ag \cite{Ag98GT-Hu}.  Hu {\em et al\/} reported the total
strength in \up98Ag to be 2.7(4), giving the hindrance factor
$h^{exp}=4.27_{-0.55}^{+0.74}$, since the calculated $B_{Q_{EC}}$ is 11.53
(this is 92\% of the total Gamow-Teller strength inside the
$Q_{EC}=8.24$ MeV window).  Thus the hindrance in two Ag isotopes,
\up97Ag and \up98Ag, are of the order of 4.25, and we adopt this value
for the total hindrance factor $h$. We did not consider heavier nuclei
for the hindrance estimate, because they are further away from our
region of interest, and the possible $Z$-dependence of this factor is
not clear \cite{Ag97GT-Hu}.

In this region, the only available total Gamow-Teller strength
measured using $(n,p)$ reaction is for \up90Zr. Raywood {\em et al\/}
\cite{Zr90GT-Raywood} deduced a value of $1.0\pm0.3$ for the total
strength.  Our calculated total strength for this isotope is
$S(GT_+)=0.34$.  We should note, however, that in our restricted model
space there is only one 1\parp\ state in \up90Y. These values can
also be compared to recently reported measurements of $3.0\pm1.9$ for
the total strength by Sakai and Yako \cite{Sakai04}.

\begin{figure}[tbp]
\centerline{\psfig{figure=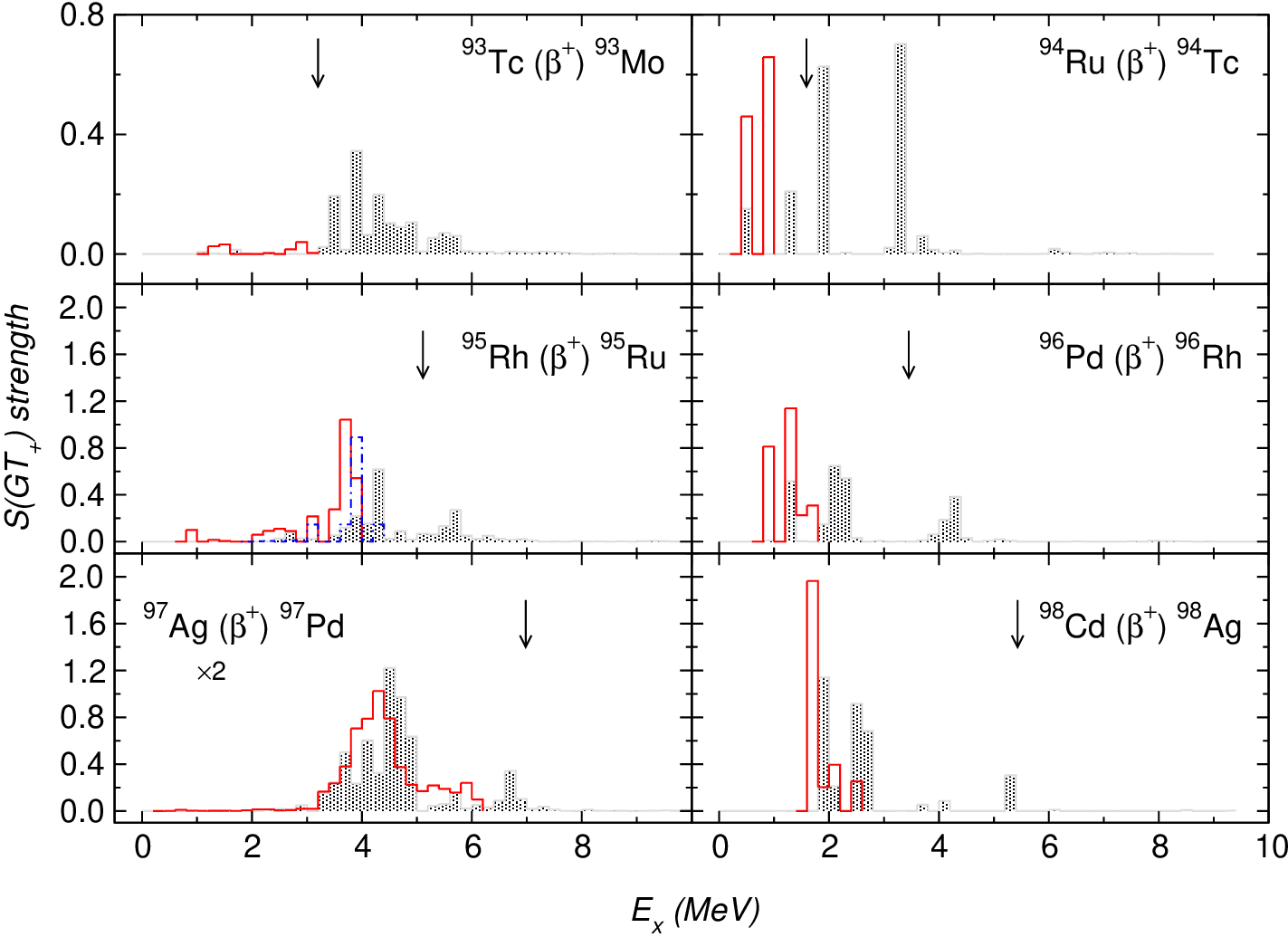,width=0.5\textwidth,angle=270}}
\caption{
\label{fig-N50GTExpDistr}\label{fig-Tc93GTcompar}\label{fig-Ru94GTcompar}\label{fig-Rh95GTcompar}\label{fig-Pd96GTcompar}\label{fig-Ag97GTcompar}\label{fig-Cd98GTcompar}
  (Color online) 
  Gamow-Teller strength distributions from the ground states of $N=50$
  isotones as a function of the excitation energy in the daughter
  nucleus.  Two experimental sets are plotted for the decay of \up95Rh
  (the values are taken from \cite{N50GT-Brown,N50GT-Johnstone}).  
  The gray-shaded
  area is the calculated strength, while the histogram represents
  experiment.  The experimental data suffer from detector sensitivity
  limits near the $Q$-limit (indicated by an arrow), with an exception
  of \up97Ag where the data were obtained using TAS
  \cite{Ag97GT-Hu}. We show the strength in \up97Ag
  scaled by a factor of 2.  The bin size is 0.2 MeV.}
\vspace{4mm}
\centerline{\psfig{figure=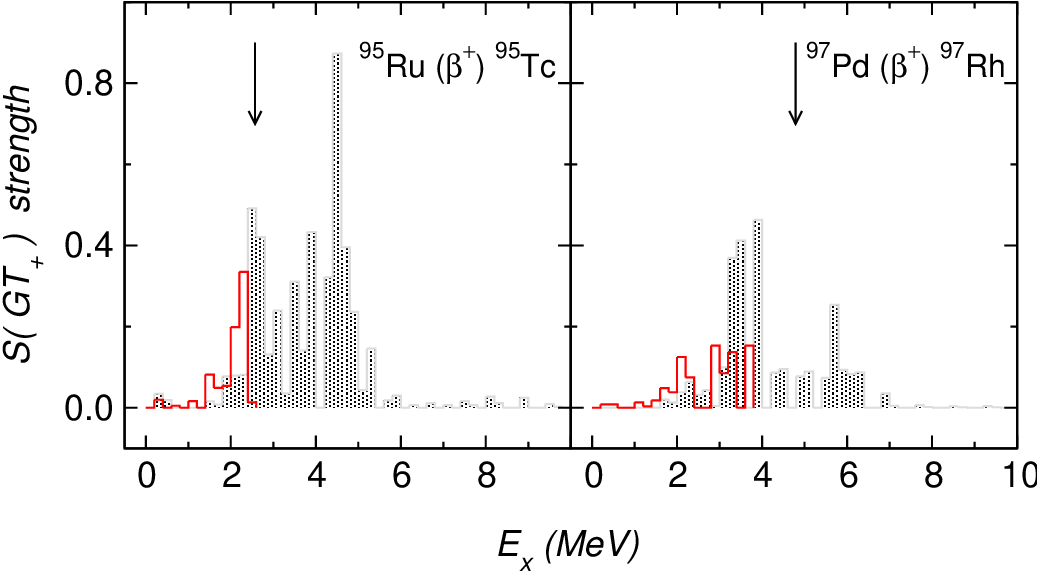,width=0.5\textwidth,angle=270}}
\caption{
  \label{fig-Ru95GTcompar} \label{fig-Pd97GTcompar}
  (Color online) 
  Gamow-Teller strength distributions from the ground states of two
  $N=51$ isotones: \up95Ru and \up97Pd (values are taken from
  \cite{N51GT-Johnstone}). See also the caption of Fig.\
  \ref{fig-N50GTExpDistr}. 
}
\vspace{4mm}
\centerline{\psfig{figure=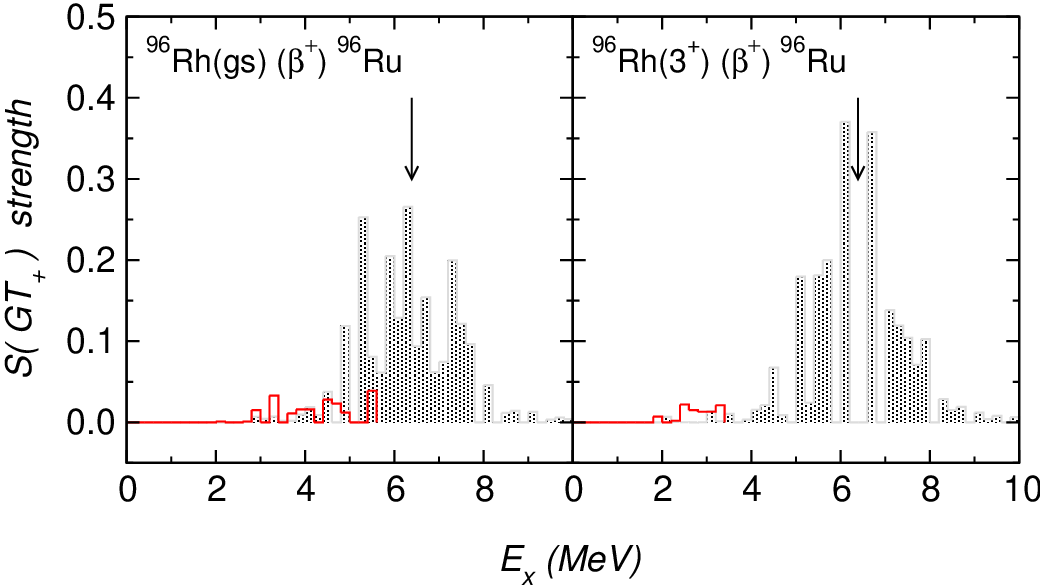,width=0.45\textwidth,angle=270}}
\caption{
\label{fig-Rh96GTcompar} \label{fig-N51GTExpDistr} 
  (Color online) 
  Gamow-Teller strength distributions
  from the ground state and first excited states of \up96Rh
  (values are taken from
  \cite{N51GT-Johnstone}).
  See also caption of Fig.\ \ref{fig-N50GTExpDistr}.
 }
\end{figure}

We turn now to strength distributions.  Our calculated Gamow-Teller
distributions in the decay of nuclear systems with a few valence
protons, like Zr or Mo isotopes, have the strength concentrated in a
narrow energy range (less than 0.5 MeV) or sometimes in only one
transition.  The strength in systems with $Z_v^p>4$ (Tc and above) is
distributed over the energy range of about 4 MeV. The Nb isotope chain
is intermediate in this respect, because in the lowest configuration
Nb has only one valence proton in the $g_{9/2}$ shell, while decays to
Zr isotopes are distributed over several states. We show these
systematics using a few examples below.

In Figs.\ \ref{fig-N50GTExpDistr}-\ref{fig-Rh96GTcompar}, we compare
the calculated Gamow-Teller distributions with available data
collected from several sources. All measurements were done using
$\gamma$-ray spectroscopy. Data for $N=50$ and 51 isotopes were
obtained from Refs.\
\cite{N50GT-Brown,N50GT-Johnstone,N51GT-Johnstone} (see also
references there in).  The $GT$ distribution in \up97Ag
\cite{Ag97GT-Hu} was obtained with a TAS measurement. We note that in
comparisons to experimental data, we do not include the sensitivity
limits of experimental detectors. This sensitivity artificially cuts
off Gamow-Teller strength near the $Q$-window so that calculated
states are often not observed even 2 MeV below the $Q$-window (see,
e.g., \cite{N50GT-Brown,Ag97GT-Hu,In102GT-Gierlik}). For this reason,
comparisons to experiment are somewhat difficult, and one should focus
on unambigous regions of low-lying strength. In Table \ref{tbl-GTQec}
we listed fractions of the calculated Gamow-Teller strength that lie
inside the $Q$-window.  A comparison to the calculation by Brown and
Rykaczewski \cite{N50GT-Brown} reveals some interaction dependence in
the values. For example, they estimated $f_{Ec}=29\%$ and 99\% in the
decays of \up94Ru and \up96Pd, while our estimates are only 19\% and
72\%, respectively. Johnstone \cite{N50GT-Johnstone,N51GT-Johnstone},
following a different approach, estimated significantly higher
fractions of the strength inside the $Q$-window for most cases except
\up95Ru.

\begin{table}[tbh]
\caption{\label{tbl-GTQec}
  Fraction of the calculated Gamow-Teller strength inside
  $Q_{EC}$ window, $f_{Ec}=B_{Q_{EC}}/GT_+$
}
\begin{tabular}{lrrrl}
\hline
Reaction\hspace{1.5cm} & \multicolumn{1}{c}{$GT_+$} &\multicolumn{1}{c}{$B_{Q_{EC}}$} &
\multicolumn{1}{r}{$f_{Ec},$} & \% \\
\hline
\up93Tc$(\beta^+)$\up93Mo & 6.12 & 0.13 & 2 \\
\up94Tc$(\beta^+)$\up94Mo & 5.74 & 0.13 & 2 \\
\up94Ru$(\beta^+)$\up94Tc & 7.89 & 1.53 & 19 \\
\up95Tc$(\beta^+)$\up95Mo & 5.43 & 0.08 & 1 \\
\up95Ru$(\beta^+)$\up95Tc & 7.49 & 0.59 & 8 \\
\up95Rh$(\beta^+)$\up95Ru & 9.41 & 6.35 & 67 \\
\up96Tc$(\beta^+)$\up96Mo & 5.36 & 0.04 & 1 \\
\up96Rh$(\beta^+)$\up96Ru & 9.05 & 5.14 & 57 \\
\up96Pd$(\beta^+)$\up96Rh & 11.18& 8.08 & 72 \\
\up97Rh$(\beta^+)$\up97Ru & 8.53 & 2.60 & 30 \\
\up97Pd$(\beta^+)$\up97Rh & 10.90& 7.38 & 68 \\
\up97Ag$(\beta^+)$\up97Pd & 12.71& 12.49& 98 \\
\up98Ag$(\beta^+)$\up98Pd & 12.47& 11.53& 92 \\
\hline
\end{tabular}
\end{table}

From Figs.\ \ref{fig-N50GTExpDistr}-\ref{fig-Rh96GTcompar} we note
that the calculated strength distributions follow the trend observed
in experiments. Most odd-$Z$ $N=50$ isotones and $N=51$ isotones have
little strength at low excitation energies ($E_x\simleq2$-3 MeV), with
the strength distributed among many states at a higher excitation
energy, some of which are above the $Q$-window. The strength in 
even-even nuclei, represented here by even-$Z$ $N=50$ isotones, is
concentrated in a few states. 

\begin{figure}[tb]
\centerline{\psfig{figure=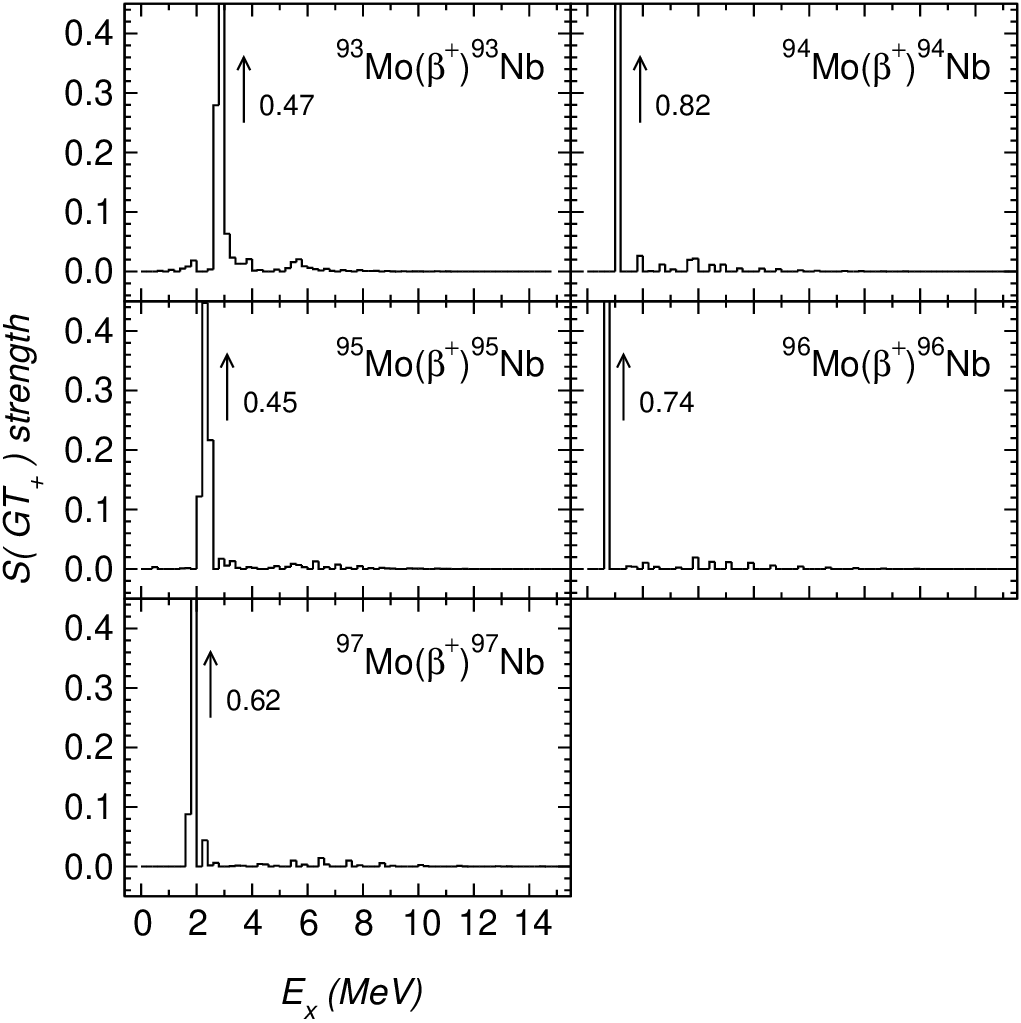,width=0.45\textwidth,angle=270}}
\caption{\label{fig-Mo9497}
  Calculated Gamow-Teller distributions in Mo isotopes with masses
  $A=93$-97 ($N=50$-55). The arrow indicates that the value inside the
  bin (the value is shown by
  a number beside it) is greater than the upper limit of the $y$-axis.
  The bin size is 0.2 MeV.
}
\vspace{4mm}

\centerline{\psfig{figure=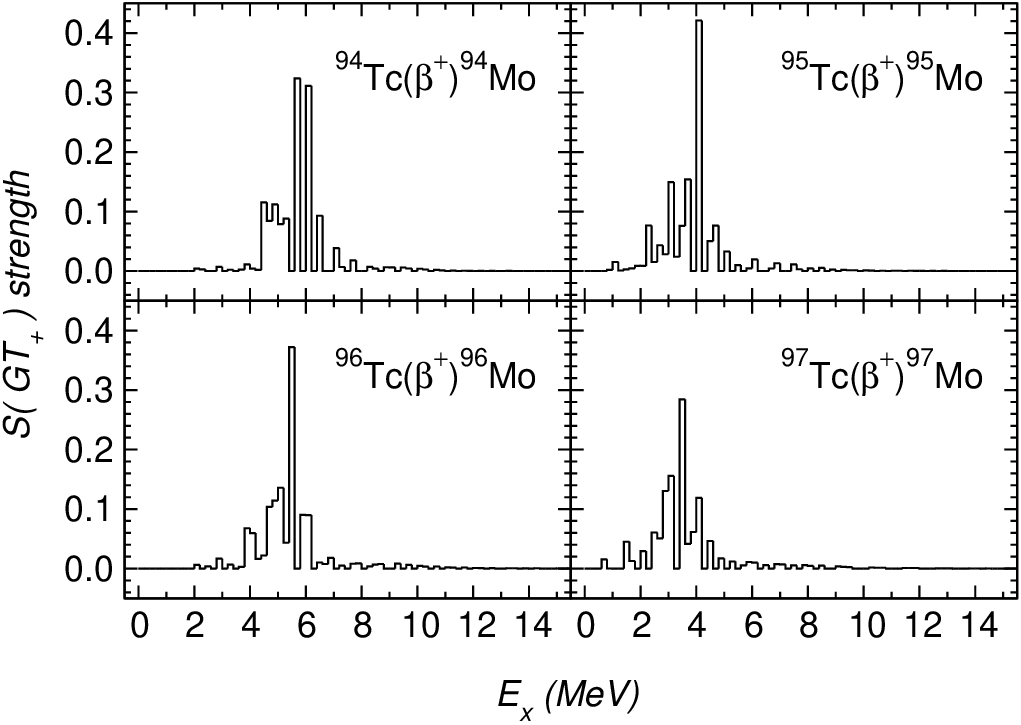,width=0.45\textwidth,angle=270}}
\caption{\label{fig-Tc9497}
  Calculated Gamow-Teller distributions in Tc isotopes with masses
  $A=94$-97. See also caption to Fig.\ \ref{fig-Mo9497}.
}
\end{figure}

The Gamow-Teller distribution in \up97Ag shown in Fig.\
\ref{fig-N50GTExpDistr} is converged (around 60 iterations per $J_f$
was required); thus the calculated shape is as good as it can be for
the interaction. The centroid of experimental Gamow-Teller strength
distribution in \up97Ag is lower than the calculation predicts:
$E_{centr}^{exp}=4.3$ MeV versus $E_{centr}^{calc}=4.7$ MeV. This is
one of the indicators that the interaction may require some
fine-tuning.

\begin{figure}[tb]
\centerline{\psfig{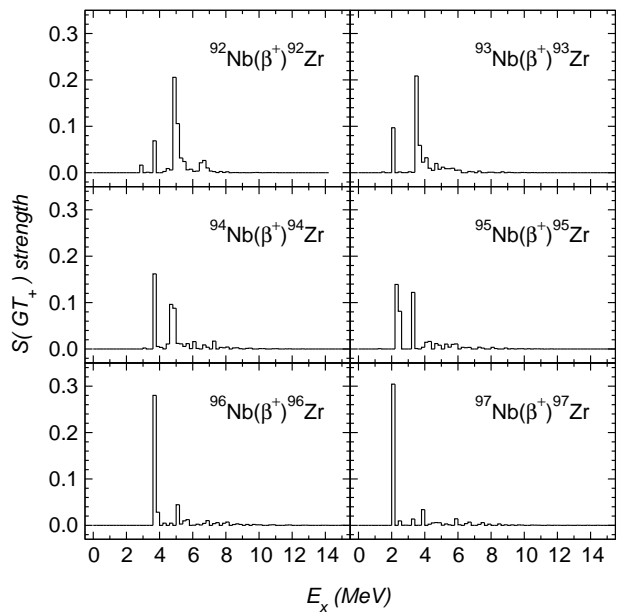}}
\caption{\label{fig-Nb9296}
  Calculated Gamow-Teller distributions in Nb isotopes with masses
  $A=92$-96. See also caption to Fig.\ \ref{fig-Mo9497}.
}
\end{figure}

We already showed and discussed parts of the calculated Gamow-Teller
distributions in Figs.\
\ref{fig-N50GTExpDistr}-\ref{fig-N51GTExpDistr}.
Having in mind upcoming experiments \cite{Mo-Zegers}
on Mo isotopes, we show the calculated distributions for the decays of
Mo isotopes with masses $A=93$-97 in Fig.\ \ref{fig-Mo9497}. The
main contributions to the total strength are located within 1 MeV
energy range around the centroid. The decays of Tc isotopes (see Fig.\
\ref{fig-Tc9497}) have the strength distributed within 4 MeV
range. These two isotope chains display the difference in the decays
of even-even and odd-odd nuclei that we discussed above.

We also show the distributions for Zr isotopes, see Fig.\
\ref{fig-Nb9296}. There we observed the migration of the strength from
higher energies to lower energies as the neutron number increases. At
\up96Zr, where the lowest configuration is a completely occupied
proton $p_{1/2}$ shell and a similar situation occurs in the neutron
$d_{5/2}$ shell, the entire strength is peaked in one transition. The
trend remains also in \up97Zr.

We turn now to a discussion of the calculated total Gamow-Teller
strength and the centroids.  The isotopic dependence of the total
strength is smooth with the strength decreasing together with the
increasing number of neutrons and/or the decreasing number of valence
protons in the $g_{9/2}$ shell.  Assuming no mass dependence, an
approximate formula can be derived: $GT_+=0.086(Z_v-1.5)(20-N_v)$.
(The factor $(20-N_v)$ is due to the relative unimportance of the
$h_{11/2}$ shell because of its negative parity.)  This form is
somewhat similar to the dependence $Z_v(20-N_v)/A$ observed in the
$pf$ shell nuclei (see e.g.\ \cite{SMMC97}). The difference may be
related to the active $j$ shells. In the $pf$ shell nuclei, discussed
in Ref.\ \cite{SMMC97}, protons predominantly occupy the $f_{7/2}$ shell;
thus its occupation is proportional to the number of valence
protons. While in our model space, the occupancy of the $g_{9/2}$ proton
shell increases due to excitations out of the $p_{1/2}$ shell via
configuration mixing. This increase is greater for isotopes closer to
the core (around 0.6), and is 0.2 for $A=97$ nuclei with
$Z_v>2$. The formula's $\chi^2$ per degree of freedom is 0.05.

\begin{figure}[tb]
\centerline{\psfig{figure=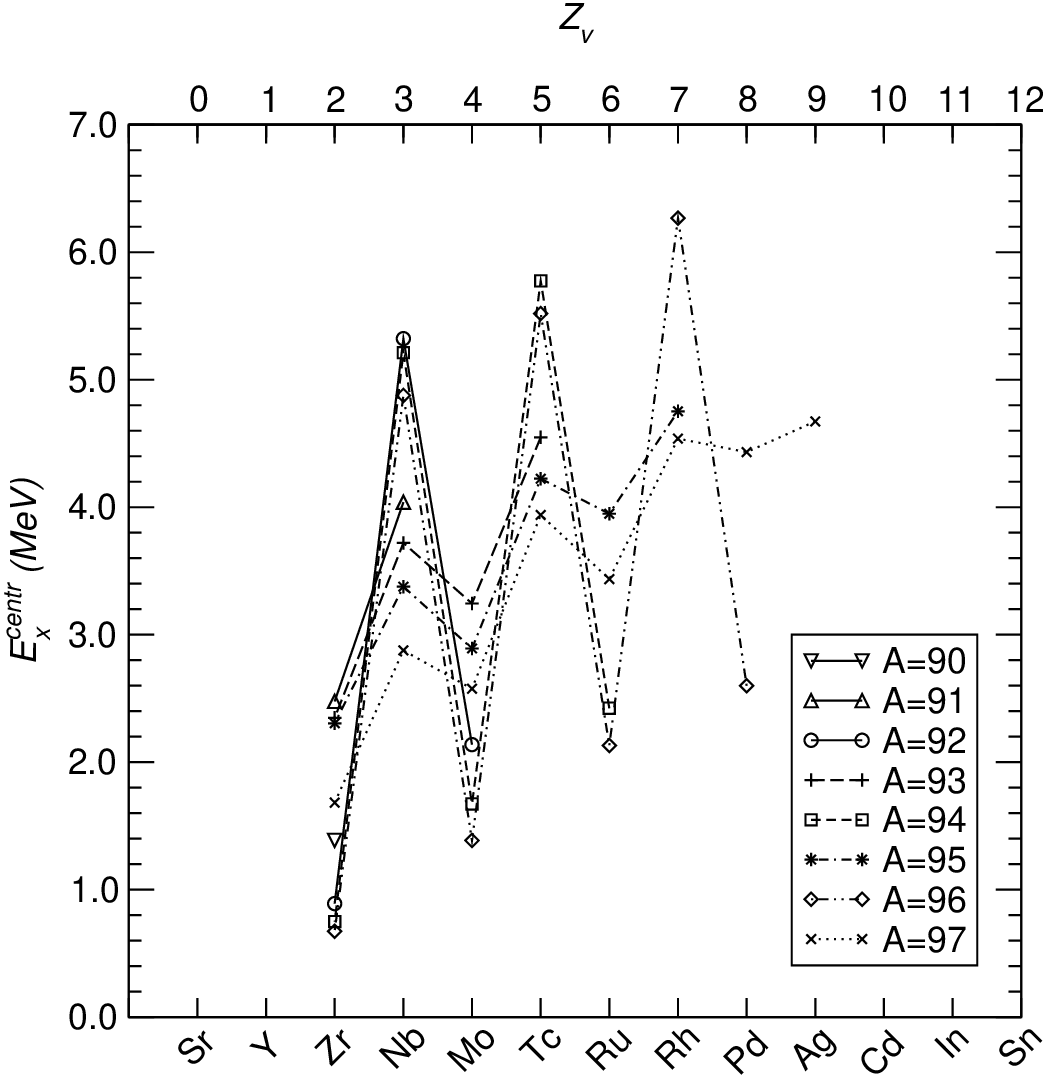,width=0.45\textwidth,angle=270}}
\caption{\label{fig-GT-CentroidEx}
  Excitation energy of the centroid in the calculated nuclei.
  For Zr$\stackrel{\tiny\beta^+}{\rightarrow}$Y decay, the excitation
  energy was calculated from the lowest positive parity state in Y.
}

\centerline{\psfig{figure=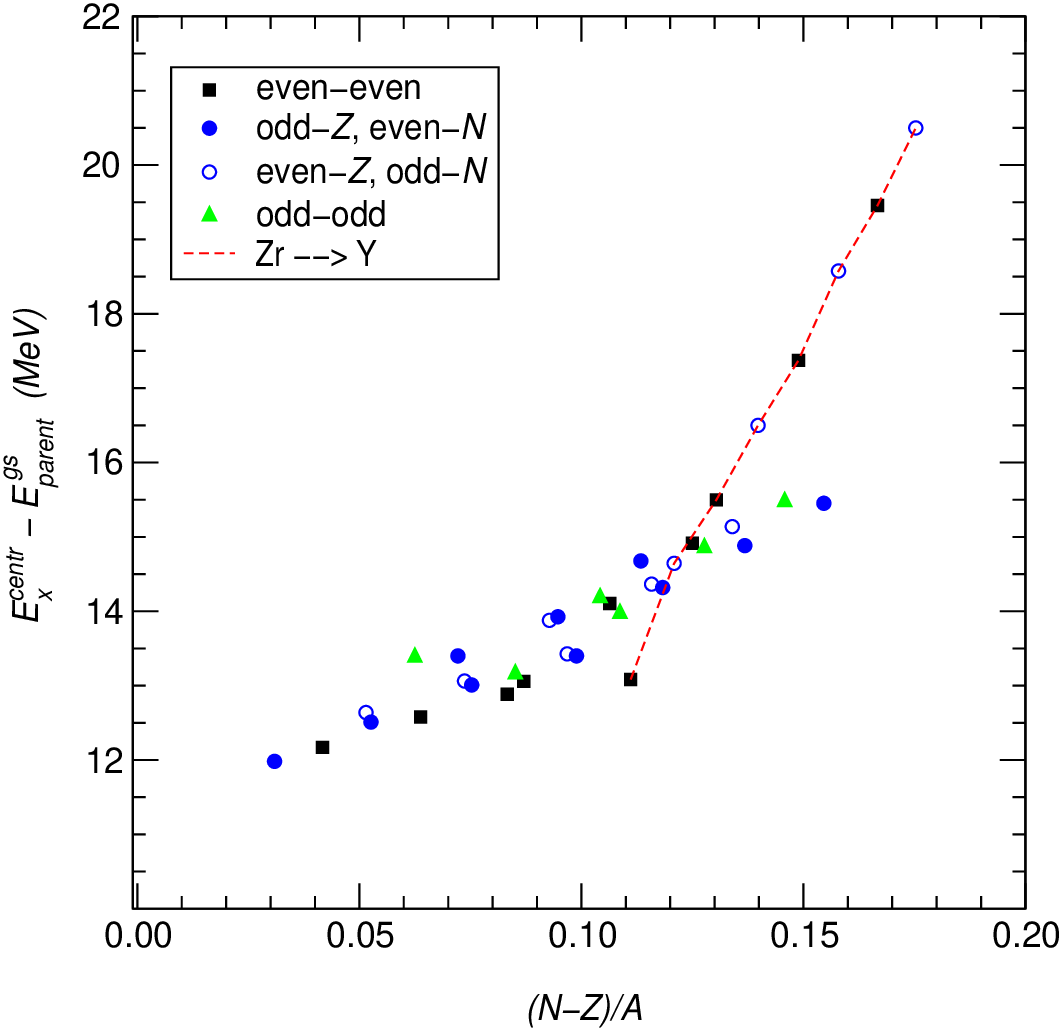,width=0.45\textwidth,angle=270}}
\caption{\label{fig-GT-CentroidEx-TzMass}
 (Color online)
  Energy of the $GT_+$ centroid with respect to the calculated ground state
  energy of the parent nucleus as
  a function of $(N-Z)/A$. See also text.
}
\end{figure}

Another systematic relates to the centroids of the $GT_+$ distribution.
If plotted with respect to the lowest positive parity state of the
daughter nucleus, the centroids of Gamow-Teller distributions show a
characteristic odd-even staggering, see Fig.\
\ref{fig-GT-CentroidEx}. They are low in even-even nuclei, high in
odd-odd nuclei, and average in odd-$A$ nuclei.  A similar trend was
observed in the mid-$pf$ shell nuclei \cite{Koonin94}.  

Langanke and
Mart{\'\i}nez-Pinedo \cite{Langanke00}
interpreted this odd-even staggering as a result of pairing
energy contributions to the mass splitting between the parent and
daughter nuclei (see also \cite{Sutaria95}).
The pairing structure goes away if the centroids are
measured with respect to the parent nucleus. We plotted the centroid
energies calculated in this way in Fig.\
\ref{fig-GT-CentroidEx-TzMass}. We also included the Coulomb energy
difference, calculated using a formula \cite{Myers66}
$E_c=0.72(Z^2/A^{1/3})(1-1.69/A^{2/3})$, but ignored the proton-neutron
mass difference and the splitting between the proton and neutron
single-particle orbits which would be present if the lowest
single-particle energies would be taken with respect to the core
nucleus, \up88Sr.  The figure shows that centroid energies indeed lose
information about the pairing structure.  It is interesting to note
that there seems to be a cross-over behavior, which we
highlighted by connecting the transitions corresponding to the decays
of Zr isotopes. These centroid energies follow a linear dependence as
well, but the inclination is different from that in other
nuclei. This behavior is probably related to the fact that $GT_+$
strength in Zr is due to proton excitations out of the $p_{1/2}$ shell.

\section{Distributed-memory shell-model code}
\label{sect-SMCode}

Our calculations were performed using a new parallel shell-model code
{\sc orpah} (Oak Ridge PArallel shell model code),
which is still under development. The basic ideas are similar to those
employed in the serial $m$-scheme computer code {\sc antoine}
\cite{Antoine,Caurier99}. However, there are differences in the
approach, since the code was developed targeting the 
distributed-memory computational paradigm. While the
distributed-memory approach
sets no limits on the available memory or the number of processors
involved in the computation, a natural
limitation occurs due to the need to communicate data from one processor to
another, a process which for collective operations scale as $N_p^2$,
where $N_p$ is the number of processors. However, even in cases when the
communication becomes unfavorable, there is still a possible trade-off
because of a greater amount of available memory.

The most time-consuming part of the shell-model problem is the operation
of the Hamiltonian on a vector to produce a new vector which 
occurs during the Lanczos procedure. This affects
load-balancing since each processor needs a similar workload for an
effective use of computational time.
We consider parallelization at two levels: the vector amplitudes 
are distributed among the processors, and each processor 
produces a portion of the final vector in a time-balanced way. 
We then redistribute the final partial vectors to the 
appropriate positions after each iteration.

Similar to {\sc antoine}, the code numerically builds ``blocks'' of
identical-particle Slater determinants having the same quantum numbers
and sets up tables allowing construction of the elements of the
Hamiltonian matrix \cite{Caurier99}. Differences arise from the
parallel implementation. If the dimension of the model space is $D$,
then the part of amplitudes which reside on a particular processor has size
$D/N_p$. For a sufficiently large number of processors, $D/N_p$ can get
smaller than the size of the largest block. This would place the limit
for the maximum reasonable number of processors. Operations involving this
block are also the most time-consuming. We decided to split the
blocks in order to have smaller pieces of tasks, which could be
distributed among the processors more efficiently. 
During the operation of the Hamiltonian acting on a vector,
some of the amplitudes are prefetched and others are requested during
the calculation. There is no predefined communication pattern, 
since some amplitudes can be delivered with some delay
while the process would still be able to employ the
ones already present in the memory. To enable this disconnection of
computation and communication, each processor consists of two threads
responsible for those two tasks. Those threads communicate via shared
variables. The need to deliver amplitudes creates a communication
overhead on top of the time needed to produce the final Lanczos
vector. The latter operation is well balanced (i.e., is inversely
proportional to the number of processors), while the overhead depends on
the number of processors involved in the calculation.  The final Lanczos
vector is reorthogonalized to previous vectors. Due to orthogonality
of the basis,
each processor can produce the partial sum of a scalar product,
and a global communication is needed only to obtain the total sum.

There is no coded-in restriction on the number of processors, with the
exception that the minimum number of processors is two, because of the
manager-worker algorithm employed in the Hamiltonian table set-up
procedure. The current version of the code can calculate eigenvalues
and eigenvectors of the Hamiltonian, the total angular momentum and
isospin, as well as $GT$ properties.  Some computations were done on a
2-4 CPU computer; others were done at the NERSC computer Seaborg using up
to 80 processors.  The largest problem that we tried to solve was the
ground state energy of \up52Fe ($D=110\times10^6$) on 48 processors,
but the current set-up did not allow us to reach such dimensions in
the region of our study.  In addition, a further improvement in the
performance is required before the code assessment is done, though the
distributed-memory computation is a venue to solve larger interacting
shell-model problems.

\section{Summary}
\label{sect-Summary}

We calculated nuclei above \up88Sr having masses $A=90$-97 using a
realistic effective interaction derived from the CD-Bonn
potential. The agreement between the calculated and measured spectra
is satisfactory. Improvements to the interaction through fine tuning
of the matrix elements could be useful to obtain the finer
spectroscopic details, including the level ordering or the placement
of negative-parity states in several nuclei, but this is beyond the
scope of this exploratory work.

We also calculated the total Gamow-Teller strength and strength 
distributions for the decay in the electron capture direction.
We found that the total strength follows the
single-particle estimate based on the $\pi g_{9/2}$ and $\nu g_{7/2}$
occupation numbers obtained from the ground state wave functions of
the parent nucleus, although the values slightly differ from a
naive single-particle shell-model picture. Calculated strength distributions
appear to reasonably recover experimental distributions in regions that
are unaffected by detector sensitivity limits. From TAS data on $^{97}$Ag,
we were able to obtain an estimate of the phenomenological quenching
factor relative to single-particle estimates. Furthermore, our 
$GT$ distribution for \up97Ag reproduces the measured data,
see Fig.\ \ref{fig-Ag97GTcompar}.  
By analyzing the centroids of the Gamow-Teller distributions, we found that
the odd-even staggering behavior disappears if the centroids are
measured from the parent ground state, as was suggested by Langanke
and Mart{\'\i}nez-Pinedo \cite{Langanke00}. We also observed that the
centroids measured in this way have 
a quasi-linear dependency on the parameterization $(N-Z)/A$,
with a different
inclination for nuclei where no protons are present in the
$g_{9/2}$ shell in the non-interacting picture.
Finally, we made several
predictions of the strength distributions for measurements that may
soon be available in mass $A=92$-97 nuclei.

The description of low-lying Gamow-Teller strength distributions in a
large range of nuclei (from roughly mass $A=50$ to mass $A=150$) is
one important ingredient in understanding type-II supernova
explosions, since electrons get captured by nuclei through these
levels.  Of course, this is not the whole story since first- and
second-forbidden transitions (typically difficult to access in the
laboratory) also play an important role in the cross section and rate
calculations relevant for supernova. For low-energy capture, the
Gamow-Teller transitions will dominate.  They also dominate the
neutrino-nucleus scattering processes that may occur at later times in
the supernova event. While theory can produce many things,
measurements are necessary to validate the estimates. Such is the case
in the nuclear region discussed in this paper. We eagerly await
experimental comparisons to our calculations.


\section*{Acknowledgments}
We are pleased to acknowledge useful discussions with
T.\ Engeland, M.\ Hjorth-Jensen, K.\ Langanke,  K.\ Rykaczewski,
R.G.T.\ Zegers, B.A.\ Brown, and I.P.\ Johnstone. We are also grateful to
M.\ Karny and L.\ Batist for providing us TAS data on \up97Ag.
Oak Ridge National Laboratory (ORNL) is managed by 
UT-Battelle, LLC, for the U.S.\  Department of Energy under
Contract No.\ DE-AC05-00OR22725. A.J.\ work was partially supported by
the Department of Energy through the Scientific Discovery through
Advanced Computing (SciDAC) program.


\end{document}